\documentstyle[aps,prl,amsmath,amssymb,multicol,citesort,epsfig]
{revtex}

\begin{document}

\title{Scaling theory for the free-energy barrier to
homogeneous nucleation of a non-critical phase
near
%an Ising-like
a critical point}
\author{{\bf Richard P. Sear}\\
~\\
Department of Physics, University of Surrey\\
Guildford, Surrey GU2 7XH, United Kingdom\\
email: r.sear@surrey.ac.uk}

\maketitle

\begin{abstract}
Homogeneous nucleation of a new phase near an Ising-like critical point of
another phase transition is studied.
A scaling analysis shows that the
free energy barrier to nucleation contains a singular term with
the same scaling as the order parameter associated with the critical point.
The total magnetisation of the nucleus scales as the
response function and so it diverges. Vapour-liquid critical points
are in the Ising universality class and so our results imply that near
such a critical point the number of molecules in a nucleus of a another
phase, such as a crystalline phase, diverges as the isothermal
compressibility.
The case where symmetry prevents coupling
between the nucleus and the order parameter is also considered.
\end{abstract}

\begin{multicols}{2}

\section{Introduction}

\begin{figure}
\begin{center}
\epsfig{file=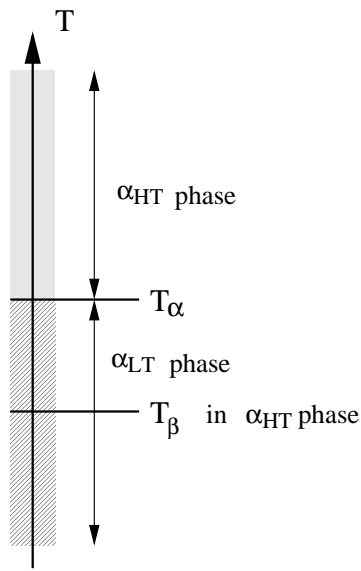,width=2.0in}
\end{center}
\caption{
\lineskip 10pt
Schematic phase behaviour along the temperature axis, at $h=0$.
The shaded (hatched) region denotes the temperature range over which
the high (low) temperature phase of transition $\alpha$
is the equilibrium phase. The temperatures of the two transitions
are labelled by $T_{\alpha}$ and $T_{\beta}$.
\label{pd}
}
\end{figure}

It is rare for a condensed matter system to possess only one phase
transition, which raises the question of how neighbouring phase
transitions affect each other. There are many
possible effects, and here we consider one of them, namely
how nucleation associated with one
phase transition is affected by the critical point of a second transition.
Consider a first-order phase transition, transition $\alpha$, which occurs
at a temperature $T_{\alpha}$. Above $T_{\alpha}$ the high-temperature
phase, call it $\alpha_{HT}$, is the equilibrium phase while below it the
low-temperature phase, $\alpha_{LT}$, is the equilibrium phase.
Fig.~\ref{pd} is a schematic of the temperature axis of the phase diagram.
If we cool the high-temperature phase below $T_{\alpha}$ then it
ceases to be the equilibrium phase. At some degree
of supercooling a nucleus of the low temperature phase will appear and grow
into a macroscopic $\alpha_{LT}$ phase. The initial step, the formation
of a nucleus of the $\alpha_{LT}$ phase, is an activated process --- a
free energy barrier must be surmounted \cite{note1}.
The rate at which nuclei overcome a barrier of height
$\Delta F^*$ scales as $\exp(-\Delta F^*/kT)$ and so this rate
is very sensitive to the barrier height \cite{debenedetti}.
Here we calculate $\Delta F^*$ for nucleation
near a critical point of another phase transition,
which we call transition $\beta$. Transition $\beta$ is Ising-like.
So, we have two phase transitions:
a strongly first order phase transition $\alpha$, whose dynamics
we are studying, and a transition with a critical point, transition
$\beta$, which lies at a temperature $T_{\beta}<T_{\alpha}$ and
so occurs within the $\alpha_{HT}$ phase when it is not the true
equilibrium phase. At an Ising critical point
the thermodynamic functions contain singular power-law terms
with exponents which depend only on dimensionality
\cite{kadanoff,chaikin}.
Below, we show that $\Delta F^*$ also
contains a singular
power-law term with exponents which depend only on dimensionality.
The singular part of $\Delta F^*$ has the
same form as that in the order parameter, so for example, it
scales with temperature as $|T-T_{\beta}|^{\beta}$, where
$\beta$ is the usual critical exponent \cite{chaikin,kadanoff}.
This singular part means that the derivative of $\Delta F^*$ with
respect to temperature diverges as the critical point is approached:
the barrier to nucleation is very sensitive to changes
in temperature near the critical point.
Which is true
for {\em any} system with an Ising-like critical point.

The scaling analysis here follows on from mean-field calculations
performed by the author \cite{sear01a,sear01b}, see also
Ref.~\cite{talanquer98}. Which in turn followed on from
the pioneering simulations of ten Wolde and Frenkel
\cite{tenwolde97} who were the first to appreciate that critical
fluctuations could reduce $\Delta F^*$ near
the critical point of a metastable fluid-fluid transition.
The results
of the mean-field calculations are consistent
with the scaling analysis here; if we take the equations of the
present work and set the exponents to their (incorrect) mean-field
values we reproduce the exponents found in Refs.~\cite{sear01a,sear01b}.
Finally, see the work of Dixit and Zukoski \cite{dixit00} for an
alternative approach to nucleation near a critical point; one
where dynamics are explicitly but approximately included.
%Our results apply to all systems with a critical point in the
%universality class of the Ising model.

Perhaps the easiest
systems in which to observe the effects found here,
are mixtures of
liquids. There nucleation of either the vapour phase or a
crystalline phase may occur near the continuation
of a line of critical solution temperatures, i.e.,
near to a critical point of liquid-liquid demixing.
A line of critical solution temperatures does not stop when the
liquid ceases to be the equilibrium phase but continues into the
region where it is metastable with respect to a transition to a vapour
or crystalline phase. It is
this continuation that affects nucleation.

The study of nucleation near a critical point was sparked by interest
in the nucleation of the crystalline phase of globular proteins
\cite{piazza00}. At least some globular proteins have a fluid-fluid
critical point within the fluid-crystal coexistence region
\cite{broide91,muschol97}.
The fact that the predictions made here
are universal is very useful with regard to protein crystallisation
as the interactions between protein molecules are generally rather
poorly understood,
particularly before they have been crystallised and their
structure determined \cite{durbin96}.
We know that our predictions will apply to nucleation near any
fluid-fluid critical point found in a protein solution, without
needing to know anything about the
protein-protein interactions.

%Ising background

\section{Theory for nuclei which couple to the order parameter}
\label{secm}

We will describe the critical point of transition $\beta$ using
the standard language of Ising spin systems, although
of course our results also apply to critical points in fluids.
The order parameter
is $m$, the temperature $t=(T-T_{\beta})/T_{\beta}$, and the field is
$h$ and couples to the
order parameter as $-hm$. The
critical point is then at $t=0$, $h=0$, see Fig.~\ref{pd}.
For fluid systems the
external field $h$ is the chemical potential minus that at
coexistence, and $m$ is a density
difference.
We set the lattice spacing to $a$.
For definiteness
we set $m$ to be positive in the nucleus. The symmetry between $+m$ and $-m$
means we can do so without loss of generality.

%\section{Nucleus in a near-critical system}

The dynamics of a first-order phase transition starts with
nucleation \cite{debenedetti,chaikin}, where a microscopic nucleus
of the new phase first appears.
The nucleus at the top of the barrier, with the highest free energy
$\Delta F^*$, determines the rate and so we will consider only this
nucleus. It is generally called the critical nucleus.
Note that conventionally the word critical
is used to denote both a continuous transition and the nucleus
at the top of the barrier although there is no connection between the
two uses of the word. Although this terminology is a little
unfortunate we will use it here and so we will be studying
a critical nucleus near a critical point.

For nucleation at a strongly first-order phase
transition (which transition $\alpha$ is assumed to be),
the nucleus must contain a compact {\em core}
of spins near their state in the bulk $\alpha_{LT}$ phase,
as only in this state do
they have a lower free energy than in the $\alpha_{HT}$ phase. For all
strongly first-order transitions, the free energy will be
higher for spins intermediate between the two phases, here
$\alpha_{HT}$ and $\alpha_{LT}$. It is this core which will grow
to form the $\alpha_{LT}$ phase.
See for example
Refs.~\cite{debenedetti,chaikin} which discuss the classical theory
for nucleation which is based on this observation.

\begin{figure}
\begin{center}
\epsfig{file=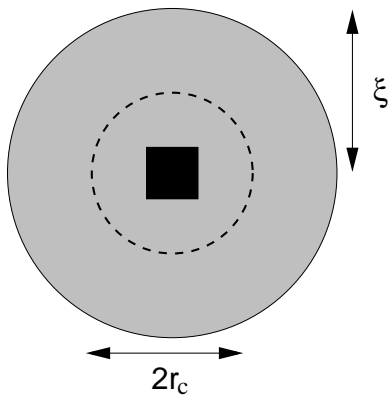,width=2.0in}
\end{center}
\caption{
\lineskip 10pt
Schematic of a nucleus of the ordered phase of transition $\alpha$
near transition $\beta$. The core of the ordered phase of transition
$\alpha$ is solid black, and the perturbation this causes in the
surroundings is the shaded circle of radius the correlation length $\xi$.
The spherical boundary with
radius $r_c$ is denoted by the dashed circle.
\label{figschem}
}
\end{figure}

Now, this core of spins will
perturb the surrounding spins out to a distance of the correlation
length $\xi$. Thus, the volume perturbed is of order $\xi^d$ and so
diverges as the critical point is approached. $d$ is the dimensionality.
The volume around the core where $m$ is small we term the {\em fringe}.
As in our earlier mean-field calculations \cite{sear01a,sear01b} we split
the nucleus into a core and a fringe. This is illustrated schematically
in Fig.~\ref{figschem}.
As $\alpha$ is a strongly first-order transition,
the order parameter of transition $\beta$, $m$, in the core
will be of order unity, i.e., not small, and therefore far above
the values of $m$ reached by critical fluctuations. The critical
fluctuations will however affect the fringe where $m$ is small.
We note that the compact core of spins
resembles a small colloidal particle in that
it is a compact, roughly spherical object, which interacts with and thus
perturbs its surroundings. The problem of a colloidal
particle in a near critical system has been considered, see
Refs.~\cite{degennes81,burkhardt95,eisenriegler95,hanke99}.

The core of the nucleus will be much smaller than the correlation
length $\xi$ near the critical point as it will be only
a few lattice sites across whereas $\xi$ diverges \cite{cntnote}.
Thus, near the critical point, the core is
a point-like
perturbation which couples to the order parameter $m$. It will couple
to $m$ unless this is prevented by a symmetry. Point-like
perturbations near a critical point are generic ---
all point-like perturbations to the order parameter $m$
perturb large length-scale critical fluctuations in the same
way, differing only in a scale factor.
Near the critical point where $\xi\gg a$, scale
invariance implies that any perturbation of size much less than $\xi$
must cause a order parameter perturbation of the same form, because
all such perturbations are simply interrelated by a change of length-scale.
The presence of a nucleus at the origin ${\bf 0}$ must therefore
produce (at large distances) the generic perturbation to $m$.
We define the thermal average excess of
$m({\bf r})$ due to the presence of a nucleus at the origin as
\begin{equation}
<m_n({\bf r})>=<m({\bf r})>_n-<m({\bf r})>,
\label{mndef}
\end{equation}
where $<m({\bf r})>_n$ ($<m({\bf r})>$)
is the thermal average value of $m({\bf r})$ when
there is (is not) a nucleus at the origin. Compare this to the 2-point
correlation function, $G({\bf r})$, for the order parameter, defined by
\begin{equation}
G({\bf r})=<m({\bf 0})m({\bf r})>-<m({\bf 0})><m({\bf r})>.
\label{2pt}
\end{equation}
Note that $G({\bf r})$ is the thermal average of $m({\bf r})$ if
$m({\bf 0})$ is fixed to 1.
As a point perturbation is generic
for the fluctuations on length scales $\gg a$, the function
$<m_n({\bf r})>$ can only differ from $G({\bf r})$ by a scale factor
when $r\gg a$. Calling this scale factor $C$,
\begin{equation}
<m_n({\bf r})>= CG({\bf r})
%r^{-(d-2+\eta)}{\cal F}_{\pm}(r/\xi_{\pm})
~~~~~r\gg a.
\label{mn}
\end{equation}
$C$ will depend on the properties of the core. This result has been
found before not for a nucleus but for a small colloidal
particle immersed in a near-critical fluid
\cite{degennes81,burkhardt95,eisenriegler95,hanke99}.
Defining the
total excess of $m$ due to the presence of the critical
nucleus as $m^*$, we have
\begin{eqnarray}
m^*=\int{\rm d}{\bf r}<m_n({\bf r})>.
\label{mstar}
\end{eqnarray}
Splitting the integration at $r_c$ which satisfies $a\ll r_c\ll \xi$,
and using Eq.~(\ref{mn}) for the integrand of the integral for $r>r_c$
we have
\begin{equation}
m^*=\int_{r<r_c}{\rm d}{\bf r}<m_n({\bf r})>
+C\int_{r>r_c}{\rm d}{\bf r}G({\bf r}).
\label{ms2}
\end{equation}
Now, the response function
$\chi$ is equal to the integral over $G({\bf r})$ \cite{chaikin,kadanoff}
\begin{equation}
\int {\rm d}{\bf r}G({\bf r})=\chi,
\label{chi1}
\end{equation}
and it contains a leading order singular part $\chi_S$ coming
from fluctuations on a length-scale
of the correlation length $\xi$, and so coming from
those parts of the integral where $r=O(\xi)$. Thus the second integral
of Eq.~(\ref{ms2}) for $m^*$ contains $\chi_S$: $m^*$ is
singular at the critical point. Near the critical
point $\chi_S$ dominates $\chi$
and so the term $C\chi_S$ in the
second integral of Eq.~(\ref{ms2}) will dominate $m^*$. Thus,
we have for the leading order singularity in $m^*$, $m^*_S$,
\begin{equation}
m^*_S=C\chi_S=
C|t|^{-\gamma}X_{\pm}(h/|t|^{\beta\delta}),
\label{ms}
\end{equation}
where $X_{\pm}$ are scaling functions, and $\gamma$, $\beta$ and
$\delta$ are critical exponents. $X_+$ is the scaling function above the
transition, $t\ge 0$, while $X_-$ is the scaling function below the
transition, $t\le 0$.
The second equality is obtained by substituting the
standard scaling expression for $\chi_S$
\cite{chaikin,kadanoff}.
Close to the critical point,
the total order parameter excess associated with the nucleus, $m^*$,
diverges with the precisely the same scaling behaviour as the
order parameter response
function, $\chi$.

Having determined the size of the nucleus, we wish to know its
free energy. We start by splitting the free energy barrier into
a nonsingular part $\Delta F_{NS}$ plus a leading order singular
part $\Delta F^*$,
\begin{equation}
\Delta F^*=\Delta F_{NS}^*+\Delta F^*_S
\label{cnt}
\end{equation}
The leading order singular part, $\Delta F^*_S$,
of $\Delta F^*$, is expected to have the usual scaling form,
of a power of $|t|$ times a function of $h/|t|^{\beta\delta}$,
\begin{equation}
\Delta F^*_S=|t|^xY_{\pm}(h/|t|^{\beta\delta}),
\label{fguess}
\end{equation}
with $x$ an as yet unknown exponent.
To determine $x$ we note that the derivative of the free energy
of the system $F$ with
respect to $h$ is,
\begin{equation}
\frac{\partial F}{\partial h}=-M,
\label{dfdh1}
\end{equation}
where $M$ is the total order parameter $m$ of the system.
Thus, as $\Delta F^*$ is the free energy with a nucleus
at the origin minus that without a nucleus at the origin, then the
derivative of $\Delta F^*$ with respect to $h$ is simply the negative
of the total
order parameter with a nucleus minus that without a nucleus \cite{note},
\begin{equation}
\frac{\partial \Delta F^*}{\partial h}=-m^*.
\label{dfdh}
\end{equation}
This is for the complete $\Delta F^*$. The leading-order singular term in
$\Delta F^*$, denoted by $\Delta F^*_S$, will yield minus the
leading-order singular term in $m^*$, upon differentiation with respect
to $h$
\begin{equation}
\frac{\partial \Delta F^*_S}{\partial h}=-m^*_S.
%=- C\chi_S=-
%C|t|^{-\gamma}X_{\pm}(h/|t|^{\beta\delta}),
\label{dfsdh}
\end{equation}
The derivative with respect to $h$ of scaling functions
like $Y_{\pm}(h/|t|^{\beta\delta})$, brings an additional $|t|^{-\beta\delta}$
factor. Thus for Eqs.~(\ref{fguess}) and (\ref{dfsdh})
to yield the correct exponent
$-\gamma$ for $m^*$ we require that $x-\beta\delta=-\gamma$. Using
the relation between exponents $\beta\delta=\beta+\gamma$
\cite{chaikin,kadanoff} we obtain $x=\beta$. So, setting
$x=\beta$ in Eq.~(\ref{fguess}) we obtain
\begin{equation}
\Delta F^*_S=|t|^{\beta}Y_{\pm}(h/|t|^{\beta\delta}),
\label{fs}
\end{equation}
the singular part of the free energy scales with the same exponent, $\beta$,
as the bulk order parameter. This is for Ising-like systems but
we expect that analogous singular terms will be found near
critical points of transitions with order parameters which are
not scalars, for example in the Heisenberg model where the
order parameter is a three-dimensional vector.

As $\beta>0$, $\Delta F^*_S=0$ at the critical point, and so
the universal singular part is zero. The free energy barrier $\Delta F^*$
at the critical point is equal to some non-universal value.
However, as
$\beta<1$, the derivative of $\Delta F^*$ with respect to temperature
diverges as $|t|^{\beta-1}$ as the critical point is approached at $h=0$.
The signs of the divergences as the critical 
point is approached from above and below are
those of $Y_{\pm}(0)$ which are unknown.
However, on physical grounds we might expect critical fluctuations
to reduce the barrier to nucleation, and this is what is found
in the mean-field theory calculations for $T>T_{\beta}$ of
Ref.~\cite{sear01b}, i.e., within mean-field theory it has been
found that $Y_+(0)>0$.
If $Y_{\pm}(0)$ are both positive, then
at fixed $h=0$, the free energy barrier will have a local minimum at
$T=T_{\beta}$.
As the derivative of $\Delta F^*_S$ diverges, near the
critical point the variation in $\Delta F^*$ is dominated by
that in $\Delta F^*_S$, and so if $\Delta F^*_S$ has a minimum
at the critical point so does $\Delta F^*$.

Along the critical isotherm,
\begin{equation}
\Delta F^*_S=\Sigma \mbox{sgn}(h)|h|^{1/\delta}~~~~~t=0,
\end{equation}
where $\Sigma$ is an amplitude and the scaling
with $h$ is fixed by the requirement that it be
the same as that of $m$.
From Eq.~(\ref{dfsdh}) we see that as $m^*_S>0$ then
the $h$ derivative of $\Delta F^*_S$ must be negative. For this
to be true we must have that $\Sigma<0$. Note that $m^*_S>0$ because
we set $m$ in the core to be $>0$, if the nucleating phase has
instead $m<0$, then in that case $\Sigma>0$. $\Sigma$ is a function
not only of the properties of the critical system but of the
nucleating phase.
At the critical point the rate of decrease of $\Delta F^*$ ($=m^*$)
diverges, so near the critical point the nucleation barrier is a rapidly
decreasing function of $h$.

\subsection{Critical amplitude ratios}

In the $h=0$ limit, the scaling functions $X_{\pm}$ of $\chi_S$,
Eq.~(\ref{ms}), simplify
\begin{eqnarray}
\chi_S&=&\Gamma_{\pm}|t|^{-\gamma}~~~h=0,
%~~~~~~\nonumber\\
%\chi_S&=&\Theta\mbox{sgn}(h)|h|^{1/\delta-1}~~~t=0,
\label{chis2}
\end{eqnarray}
where $\Gamma_{\pm}=X_{\pm}(0)$.
$\Gamma_{\pm}$ are critical amplitudes, and are
non-universal. However, the critical amplitude ratio
$\Gamma_+/\Gamma_-$ is universal \cite{chaikin,guida98}.
The fact that not only the exponents but the critical amplitude ratios
are universal means that we can even make a universal prediction about the
ratio of the size of the nucleus just above the critical point to that
just below. If we take the ratio $m^*_S(+|t|)/m^*_S(-|t|)$,
we see, using Eq.~(\ref{chis2}), that
\begin{equation}
\frac{m^*_S(+|t|)}{m^*_S(-|t|)}=\frac{\Gamma_+}{\Gamma_-}
~~~~~h=0,
\end{equation}
which is universal.
As near the critical point $m^*_S$ dominates $m^*$, we have that near
the critical point: 1) the excess magnetisation associated with the nucleus
is the {\em same} in each of the two coexisting phases at some $t<0$, and 2)
that the ratio of $m^*$ at $|t|$ to the common value of the excess $m$
in the coexisting phases at $-|t|$ is universal. In three dimensions
$\Gamma_+/\Gamma_-=4.9$ \cite{chaikin,guida98}
so the nucleus is approximately
five times bigger $|t|$ above the transition than
$-|t|$ below. Via the nucleation theorem
\cite{kashchiev82,viisanen93,bowles00,bowles01} it is
possible to estimate the size of the critical nucleus.
Thus measurements
on nucleation rates can potentially be used to estimate not only
critical exponents but critical amplitude ratios.

\section{Dynamic scaling for the equilibration time near $T_{\beta}$}
\label{sec:dyn}

Now, the singular part of the barrier, $\Delta F^*_S$,
is universal because it originates in long wavelength
fluctuations, fluctuations with wavelengths $\gg a$, where the
microscopic details of the system are irrelevant. However, these
long wavelength fluctuations are slow, firstly as a direct
consequence of them having a long wavelength (this applies if $m$ is a
conserved variable \cite{chaikin}), and secondly because
as the critical point is
approached the free energy becomes a very flat function of the
order parameter which reduces the driving force which relaxes fluctuations.
Dynamic scaling theory yields for the characteristic relaxation time
$\tau_R$ (=$2\pi$ over the characteristic frequency) of a small
wavevector mode, $k<<a^{-1}$, near a critical point
\cite{chaikin,hohenberg77}
\begin{equation}
\tau_R=k^{-z}D_{\pm}(k\xi,h/|t|^{\beta\delta}),
\label{taur}
\end{equation}
the relaxation time of a mode of wavevector $k$ at temperature $t$ and
field $h$. The exponent $z$ is a dynamic scaling exponent. Dynamic
scaling exponents depend not only on the symmetry of the order parameter
and the dimensionality as the static exponents, $\beta$, $\gamma$, etc. do,
but also on whether the order parameter $m$ is conserved or not.
If $m$ is the magnetisation then it can vary locally without transport of
$m$ and so it is non-conserved, but if $m$ is a density difference then
as the total number of molecules is conserved it cannot vary in a volume
of space without transport in or out of that volume, so it is then a
conserved variable. In either case $z$ is positive.
Assuming that fluctuations with wavelengths larger than
the correlation length can be ignored we have that at a given $t$ and $h$,
the longest relevant relaxation time is that of a mode of wavevector
$k=2\pi/\xi$. Calling the relaxation time of this mode $\tau_{\xi}$
we have, from Eq.~(\ref{taur}),
\begin{eqnarray}
\tau_{\xi} &=& \xi^z(2\pi)^{-z}D_{\pm}(2\pi,0)\nonumber\\
\tau_{\xi} &=& \tau_{\pm}|t|^{-z\nu}~~~~~~h=0,
\label{tauxi}
\end{eqnarray}
where we have set $h=0$ for simplicity, the generalisation to non-zero
fields is easy. The second line used the scaling for the correlation
length $\xi=\xi_{\pm}|t|^{-\nu}$ for $h=0$, and the
amplitudes $\tau_{\pm}=\xi_{\pm}^z(2\pi)^{-z}D_{\pm}(2\pi,0)$.

Our expression for the free energy barrier to nucleation,
Eq.~(\ref{cnt}), includes a singular part $\Delta F^*_S$
coming from long-wavelength fluctuations. 
This singular part is an equilibrium quantity: it is the free energy
change of equilibrium fluctuations due to the nucleus. Note that here by
equilibrium we mean that the fluctuations in $m$ etc. have relaxed
to equilibrium, the system is not at true equilibrium as it is in the
$\alpha_{HT}$ phase below $T_{\alpha}$. Thus, our expression for
the free energy barrier, Eq.~(\ref{cnt}), is only meaningful once a time
much larger than $\tau_{\xi}$ has elapsed at that temperature and so
the fluctuations in $m$ have equilibrated.

However, $\tau_{\xi}$ diverges as the critical point is approached so the
nearer we are to the critical point the longer we have to wait before
Eq.~(\ref{cnt}) gives the correct barrier to nucleation. For $m$ conserved,
$z=4-\eta$ and so putting in the values for $\eta$ and $\nu$ we have
that $\tau_{\xi}$ scales as $|t|^{-2.5}$: it diverges rather rapidly
as the critical point is approached. For $m$ not conserved,
$z$ is close to 2 and so putting in the values for $\eta$ and $\nu$ we have
that $\tau_{\xi}$ scales as $|t|^{-1.3}$: it still diverges but
more slowly. So, the closer the critical point is approached the longer
is the waiting time before our expressions for the nucleation barrier,
and size of the nucleus are valid. This, together with the effect of
gravity on long wavelength fluctuations will limit how closely the
critical point can be approached and the behaviour predicted
by Eqs.~(\ref{ms}) and (\ref{cnt}) observed.

\section{Nuclei which do not couple to the order parameter and
non-Ising universality classes}

Section \ref{secm} considered the case of a nucleus near an Ising-like
transition, and in which the nucleus couples to the order parameter.
In this section we start by considering nuclei near an Ising-like critical
point which do not couple to the order parameter, and then move on to
consider critical points in other universality classes.
There are results for colloidal particles in fluids near Ising-like
critical points where the particles
do not couple to the order parameter \cite{burkhardt95}.
This is so when the two phases which form below $T_{\beta}$ are
related by symmetry and the nucleus does not discriminate between these
two phases. Now, whether or not a nucleus (or colloidal particle)
couples to the order parameter, it will in general couple to the energy
density $u$. The perturbation to the energy density around a particle/nucleus
is given by the energy correlation function. A scaling argument
analogous to that in the previous section can be constructed to yield the
result that the coupling to the energy density yields a
singular part in the free-energy barrier which
scales as the energy density $u$, i.e.,
\begin{equation}
\Delta F^*_S=|t|^{1-\alpha}Z_{\pm}(h/|t|^{\beta\delta}),
\label{fsu}
\end{equation}
where $\alpha$ is the specific-heat exponent, and $Z_{\pm}$ are
scaling functions. The right-hand side of Eq.~(\ref{fsu}) is the
appropriate scaling for the energy density.
Now, $1-\alpha>\beta$, and so if a nucleus couples
to both $m$ and $u$ then the singular term from the $m$ coupling,
Eq.~(\ref{fs}), will dominate and we can neglect the singularity
of Eq.~(\ref{fsu}): we did this in the previous section.
In general, we always expect the nucleus
to couple to the energy density but the resulting singularity,
Eq.~(\ref{fsu}) will be dominated by that from the $m$ coupling unless
the latter is absent in which case Eq.~(\ref{fsu}) should give the
dominant singularity in the free energy barrier to nucleation.
Note that the temperature
derivative of the nucleation barrier $\Delta F^*/T$.
varies as the heat capacity, i.e., as $|t|^{-\alpha}$
at $h=0$.
%Also, whether or not the nucleus couples to $m$ the temperature
%derivative gives the negative of the excess entropy of
%the nucleus, see Ref.~\cite{bowles01}.

So, for a nucleus which does not couple to the order parameter,
we still have a singularity in the free-energy barrier, Eq.~(\ref{fsu}),
but it is weak. In three dimensions, $\alpha=0.11$ and so the
temperature derivative of the free-energy barrier diverges only
as $|t|^{-0.11}$ at $h=0$: this weak divergence may be difficult to
observe. Also, at $h=0$, if the nucleus does not couple to $m$, then
$m^*=0$ at $h=0$ by symmetry. Thus, Eq.~(\ref{dfdh}), the field derivative
of $\partial\Delta F^*/\partial h=0$ at $h=0$.

As mentioned in the introduction, in fluid systems like
protein solutions and liquid mixtures, there is no
symmetry available to prevent the coupling of the nucleus to $m$.
So the results of the previous section apply there, and the coupling
to the energy density is subdominant.
However, let us consider systems in
universality classes other than that of the Ising model.
We speculate that our finding of
singular terms in the free-energy barrier to nucleation near a critical
point is not restricted to systems with a scalar order parameter.
Even if the order parameter is, say a two-dimensional unit vector
as it is in the universality class of the XY-model \cite{chaikin,kadanoff},
then the nucleus of a new phase should couple to the critical fluctuations.
For example, the fluid to superfluid transition in liquid $^4$He is
in the XY universality class. Now, the vector order parameter is the
phase of a wavefunction and a nucleus of another phase, e.g., the
vapour phase, will not couple to the phase of a quantum mechanical
wavefunction: $\Delta F^*$ does not depend on the local phase
of the wavefunction. However, there is no reason for the nucleus
not to couple to energy fluctuations near the fluid to superfluid
transition. If it does so then $\Delta F^*$ will contain a singular
term of the form of Eq.~(\ref{fsu}).

The fluid to superfluid transition
crosses the liquid-vapour transition
\cite{chaikin}, and so
nucleation of the vapour phase can occur near the continuation of the
superfluid-fluid transition into conditions of pressure and
temperature such that the vapour phase is the equilibrium
phase.
However, for the XY model in three dimensions, $\alpha$ is (very
small but)
negative which means the heat capacity, and thus the temperature
derivative of $\Delta F^*$, has a cusp only: it does not diverge.
This may make proving or disproving our speculation difficult.

\section{Conclusion}

We have considered the effect of a critical point on the nucleation
of a new phase of another phase transition. The order parameter
associated with the critical point is a scalar, putting the transition
in the universality class of the Ising model.
The critical
point results in power-law singularities in both the
free energy barrier to nucleation, $\Delta F^*$, and the size
of the nucleus, $m^*$. The singular term in $\Delta F^*$ has the
same scaling as the order parameter, and that in $m^*$ has the scaling
of the response function of the order parameter. Thus, for example,
as the critical point
is approached at zero field, the singular term in $\Delta F^*$ varies as
$|t|^{\beta}$, where $\beta$ is the usual critical exponent and
$t$ is the temperature minus that at the critical point.
As $\beta>0$, the
singular term in the free energy barrier tends to zero as the critical
point is approached. However, derivatives of $\Delta F^*$, both with respect
to the temperature and to the external field, diverge as the
critical point is approached: the free-energy barrier to nucleation
varies rapidly near the critical point.
These predictions are universal, they
apply to nucleation in any system in which nucleation of a new phase
occurs near a critical point in the universality class of the Ising model.

All the above holds given only that the core of the nucleus, where
the nucleus resembles the new bulk phase that is nucleating, couples
to the order parameter. It will do so unless prevented by a symmetry.
This symmetry is clearly absent in fluid systems, in which the order
parameter is a density and/or composition difference. In the previous
section we speculated that there are also singular terms in $\Delta F^*$
near critical points in other universality classes. In other universality
classes it is clear that symmetry can prevent coupling to the order
parameter, in which case the singularity in $\Delta F^*$ will be weaker.

The relaxation time diverges as the critical point is approached,
see section \ref{sec:dyn}. Thus, in experiment there is a limit to
how close the critical point can be approached before the relaxation
time becomes prohibitively long. This may limit the minimum value of $|t|$
that is achievable or it may be limited by some other factor.
For example, the protein lysozyme
has a fluid-fluid transition with a critical point within the fluid-crystal
coexistence region \cite{muschol97}, and thus is a candidate for
studying nucleation of a non-critical phase, here a crystalline phase,
near an Ising-like critical point. Far from the critical point the
characteristic relaxation time can be estimated
as the time a lysozyme molecules takes to diffuse its own diameter.
The single
particle diffusion constant of lysozyme $D=O(10^{-10}\mbox{m}^2\mbox{s})$
\cite{muschol95}, and its diameter is approximately 4nm \cite{muschol95}.
This gives a relaxation time of order $10^{-7}$s. Using this as the
order of magnitude of the amplitudes $\tau_{\pm}$, we have that at $h=0$,
the relaxation time $\tau_{\xi}$ is of order $10^{-7}|t|^{-2.5}$s
at a temperature $t$. The exponent of $t$ is $z\nu$, with $z$ taking
its value for a conserved order parameter. Now, assuming careful
temperature control the critical temperature can be approached to within,
say, $0.1^{\circ}$C. As the effective interactions between the molecules
in solution vary strongly with temperature $0.1^{\circ}$C should
correspond to $|t|=O(10^{-2}-10^{-3})$, not $|t|=4\times 10^{-4}$
as it would for temperature independent interactions. At $|t|=10^{-3}$,
we have a relaxation time $\tau_{\xi}=O(1\mbox{s})$: still rather short.
Thus, in experiments on globular protein molecules like lysozyme,
limitations on how well the temperature can be controlled are the limiting
factor; the relaxation time $\tau_{\xi}$ is always well within the range
accessible in experiment.

%Now, assuming that experiments
%become impractical for relaxation times greater than an hour, we have that
%experiments are only practical to within a temperature difference of no less
%that $|t|=10^{-4}$. This is very small, the correlation length $\xi$ and
%hence the spatial extent of the nucleus will be of order 1$\mu$m this
%close to the critical point. Also, note that this close to the critical
%point the small gradient in chemical potential due to gravity is
%significant. The lengthscale over which the chemical potential changes
%by of order $kT$ is $kT/m_Bg=O(10^{-2}\mbox{m})$, where $m_B$ is the
%buoyant mass of a protein molecule, of order $10^{-24}\mbox{kg}$,
%and $g$ is the acceleration due to gravity. Now, when $|t|$ is small
%we require, see any of the scaling functions, that $|h|/|t|^{1.6}\ll 1$,
%in order for the field $h$ to have a negligible effect. When
%$|t|=10^{-4}$ this implies $|h|\ll 10^{-6}$, which is only true over
%lengthscales much less than $10^{-4}$m. Thus in a sample of height 1cm,
%only a portion of height $0.1$mm will be as close as $|t|=10^{-4}$
%and $h=0$, to the critical point. Of course, our considerations of the
%effect of the diverging relaxation time and gravity only provide
%lower bounds to how closely the critical point of lysozyme may be approached.
%As $|t|=10^{-4}$ is a very small temperature difference it is likely that
%experimental problems of temperature control, etc., will be the limiting
%factor.

It is a pleasure to acknowledge discussions with R. Evans, A. Parry and
P. Upton.
This work was supported by EPSRC (GR/N36981).

\end{multicols}

\begin{thebibliography}{99}
\bibitem{note1} Here we consider only homogeneous nucleation, nucleation
in the bulk, far from any interface. Heterogeneous nucleation, nucleation
at an interface, either with a wall or an impurity, is in fact more
common, and is not always an activated process.
\bibitem{cntnote} Classical nucleation theory predicts that for
reasonable values of the interfacial tension
(here that between the $\alpha_{HT}$ and $\alpha_{LT}$ phases)
the $\Delta F^*$ will be of the
right order for nucleation to occur when the nucleus contains
$O(10)$ spins. As critical fluctuations change $\Delta F^*$ by an
amount $O(kT)$, this conclusion still holds near a critical point.
\bibitem{note} See Ref.~\cite{bowles01} for a careful, thermodynamic,
derivation of Eq.~(\ref{dfdh}). It is Eq.~(24) of this reference. Note
that they use language appropriate to fluids, so our $m^*$ is their
$\Delta N$ the excess number of molecules in the nucleus and the
field is not $h$ but $\mu$ the chemical potential of the molecules.
\end{thebibliography}
\end{document}